\documentclass[pra,aps,reprint,showpacs,amsmath]{revtex4-1}
\usepackage{bm}
\usepackage{graphicx}
\usepackage{dsfont}

\newcommand{\bra}[1]{\langle #1 | \,}
\newcommand{\ket}[1]{\, | #1 \rangle}

\newcommand{\expv}[1]{\langle #1 \rangle}
\newcommand{\bds}[1]{\boldsymbol{#1}}
\newcommand{\lra}{\leftrightarrow}
\newcommand{\om}{\omega}
\newcommand{\Om}{\Omega}
\newcommand{\ga}{\gamma}
\newcommand{\Ga}{\Gamma}
\newcommand{\de}{\delta}
\newcommand{\De}{\Delta}

\newcommand{\eps}{\epsilon}

\newcommand{\hlf}{\frac{1}{2}}
\newcommand{\mc}[1]{\mathcal{#1}}
\newcommand{\sig}{\hat{\sigma}}

\begin{document}

\title{Dipolar exchange induced transparency with Rydberg atoms}

\author{David Petrosyan}
\affiliation{Institute of Electronic Structure and Laser, FORTH,
GR-71110 Heraklion, Crete, Greece}


\date{\today}

\begin{abstract}
A three-level atomic medium can be made transparent to a resonant
probe field in the presence of a strong control field acting on
an adjacent atomic transition to a long-lived state, which can be 
represented by a highly excited Rydberg state. The long-range
interactions between the Rydberg state atoms then translate 
into strong, non-local, dispersive or absorptive interactions 
between the probe photons, which can be used to achieve 
deterministic quantum logic gates and single photon sources. 
Here we show that long-range dipole-dipole exchange interaction 
with one or more spins -- two-level systems represented by atoms 
in suitable Rydberg states -- can play the role of control field 
for the optically-dense medium of atoms. This induces transparency 
of the medium for a number of probe photons $n_p$ not exceeding 
the number of spins $n_s$, while all the excess photons are resonantly 
absorbed upon propagation. 
In the most practical case of a single spin atom prepared in the Rydberg 
state, the medium is thus transparent only to a single input probe photon. 
For larger number of spins $n_s$, all $n_p \leq n_s$ photon components 
of the probe field would experience transparency but with an $n_p$-dependent 
group velocity. 
\end{abstract}

\pacs{42.50.Gy, 
32.80.Ee, 
03.67.Lx, 
}


\maketitle

\section{Introduction.}
Atoms excited to the Rydberg states with high principal quantum
numbers $n \gg 1$ have very long natural lifetimes $\tau \propto n^3$ 
and strong electric dipole moments $\wp \propto n^2$ for the microwave 
transitions between the neighboring states \cite{RydAtoms}. 
The resulting long-range, resonant (exchange or F\"orster) and 
nonresonant (dispersive or van der Waals) dipole-dipole interactions 
between the atoms can suppress multiple Rydberg excitations within 
a certain blockade distance \cite{Jaksch2000,Lukin2001,rydQIrev,rydDBrev}. 
Dipole-dipole exchange interactions can mediate long-range binding 
potentials between Rydberg atoms \cite{Kiffner2012,Kiffner2013,Petrosyan2014}
and can be used to study coherent \cite{Bettelli2013,Barredo2015,Yu2016} and 
incoherent \cite{Whitlock2013,Whitlock2015} excitation 
transfer processes.

An optically-dense atomic medium can be made transparent to a resonant 
probe field whose photonic excitations are coherently mapped onto the 
atomic excitations, forming the so-called dark-state polaritons 
\cite{Fleischhauer2000,EITrev2005}. This effect is called 
electromagnetically induced transparency (EIT) \cite{EITrev2005}, 
and it is usually mediated by a control laser field driving 
the atoms on the transition adjacent to the probe resonance. 
Alternatively, the driving laser can be replaced by an electromagnetic 
mode of a resonator strongly coupled to the corresponding atomic transition
\cite{Field1993,Tanji-Suzuki2011}. For an initially empty cavity, 
the resulting vacuum induced transparency (VIT) is sensitive 
to the number of photons in the input probe pulse and can therefore 
serve as a photon-number filter \cite{Nikoghosyan2010}.

\begin{figure}[t]
\centerline{\includegraphics[width=9cm]{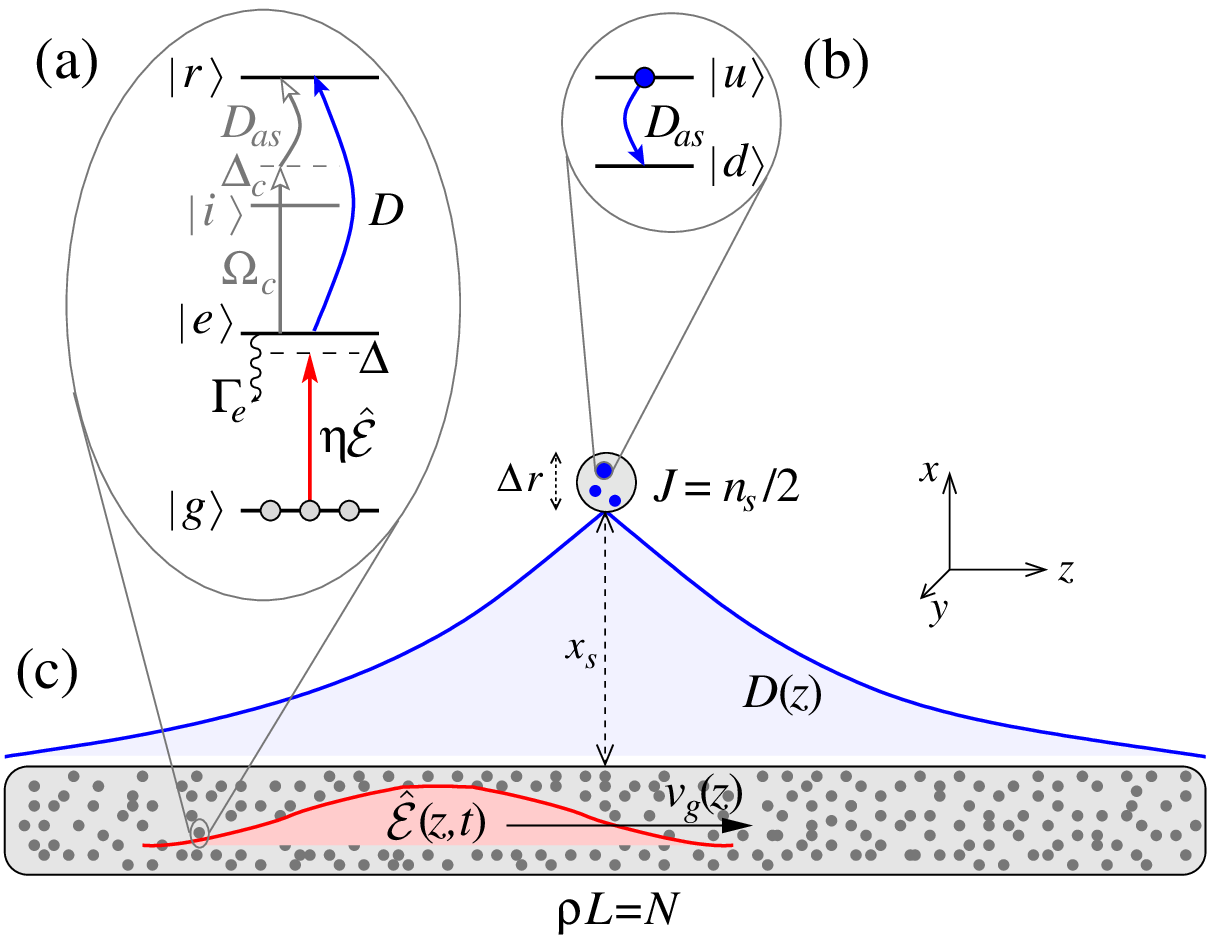}}
\caption{
Schematics of the system. 
(a) Level configuration of atoms interacting with 
the probe field $\hat{\mc{E}}$ on the transition between 
the ground state $\ket{g}$ and excited state $\ket{e}$ 
which decays with rate $\Ga_e$, while the coupling $D$ 
to the Rydberg state $\ket{r}$ is mediated by the dipole-dipole 
exchange interaction $D_{as}$ with the effective spin-$J$ 
(atoms in (b)) and an auxiliary laser $\Om_c$ detuned from 
the non-resonant intermediate Rydberg state $\ket{i}$ by 
$\De_c \gg \Om_c,D_{as}$.
(b)~The $n_s \geq 1$ atoms with the Rydberg states 
$\ket{u}$ and $\ket{d}$, confined in a small volume of size $\De r$,
form an effective spin $J = n_s/2$ which interacts 
with the medium atoms in (a) via the dipole-dipole exchange. 
(c)~The probe pulse $\hat{\mc{E}}(z,t)$ propagates with 
group velocity $v_g(z)$ along the $z$ axis in an optically 
dense atomic medium of linear density $\rho$ and length $L$. 
The atoms at positions $z$ are subject to dipole-dipole 
interaction $D(z) \equiv D(z\bds{e}_z -\bds{r}_s)$ with the effective 
spin $J$ at position $\bds{r}_s$ resulting in DEIT for the probe field.}
\label{fig:scheme}
\end{figure}

Here we propose a hitherto unexplored mechanism to attain transparency 
for a weak resonant probe field propagating in an ensemble of atoms
whose adjacent transition is strongly coupled by dipole-dipole exchange 
interaction to one or more spins -- two-level systems -- playing 
the role of a quantized control field.
In analogy with EIT and VIT, we call this mechanism 
dipolar exchange induced transparency -- DEIT. 
By employing resonant dipole-dipole interaction between 
suitable pairs of highly-excited Rydberg states, we ensure  
that the atoms of the medium are subject to a strong and long-range
dipolar exchange field of the effective spins, see Fig.~\ref{fig:scheme}. 
Each probe photon propagating in the DEIT medium with a slow group 
velocity creates an accompanying Rydberg excitation by flipping one spin. 
The number of spins $n_s$ then 
determines the maximal number of probe photons $n_p \leq n_s$ 
that can simultaneously be accommodated in the medium without absorption. 
Once all the spins are flipped, the excess $(n_p - n_s)$ photons see 
resonant two-level atomic (TLA) medium. If the medium is optically thick,
it absorbs all of the excess photons. The system can thus serve 
as a photon-number filter, with the number of appropriately 
prepared spins $n_s =0,1,\ldots$ being the switch.

We note related but conceptually different studies of Rydberg EIT 
with atoms in a ladder configuration of levels 
\cite{Friedler2005,He2011,Parigi2012,Tiarks2016,Paredes2014,Gorshkov2011,%
Pritchard2010,Petrosyan2011,Ates2011,Sevincli2011,Peyronel2012,photboundsts,%
Hofmann2013,Kuzmich13,Otterbach2013,Distante2016,Li2014,Maxwell2013,%
sphotswDuerr2014,sphotswHofferberth2014,PohlRev2016,FirstenbergRev2016}.
These schemes employ essentially conventional EIT for the probe field 
acting on the atomic transition between the ground $\ket{g}$ and intermediate 
excited $\ket{e}$ states with a classical driving field coupling state 
$\ket{e}$ to a high-lying Rydberg state $\ket{r}$. In such a medium, the probe 
photons turn into dark-state polaritons having large admixture of atomic 
Rydberg excitations. The interactions between the Rydberg-state atoms then 
lead to strong, non-local interactions between the photons. In particular, 
Rydberg mediated interactions can result in large conditional phase shifts 
\cite{Friedler2005,He2011,Parigi2012,Tiarks2016,Paredes2014,Gorshkov2011}, 
or even more dramatically, destruction of transparency of the medium within 
a blockade distance of $d_{\mathrm{b}}$ around a single propagating or stored 
Rydberg polariton \cite{Gorshkov2011,Peyronel2012,Li2014,Maxwell2013,%
sphotswDuerr2014,sphotswHofferberth2014}. 
But complete scattering of light induced by a single Rydberg excitation 
requires large optical depth per blockade distance 
$d_{\mathrm{b}} \lesssim 10\:\mu$m, which entails problems: 
Increasing the atom density and/or choosing higher Rydberg states 
$n \gtrsim 100$ to increase the range of the van der Waals  
interactions leads to strong decoherence of the Rydberg-state 
electrons \cite{TPfau20134} and thereby EIT inhibition. 

In the present scheme, the situation is in a sense reversed: 
A single spin -- atom in a suitable Rydberg state with large transition 
dipole moment -- does induce transparency of the medium for a single 
probe photon within the interaction distance $d_{\mathrm{t}}$. 
Due to the longer range of the resonant dipole-dipole interaction, 
as compared to the van der Waals interaction, DEIT with atoms 
excited to lower Rydberg states with $n \sim 80$ can extend over 
the medium of length $L = 2 d_{\mathrm{t}} \simeq 25\:\mu$m. 
The excess photons are then completely scattered upon propagation 
in the medium with moderate atom density but large optical depth.

\section{Mathematical formalism}

\subsection{The Hamiltonian of the system}

We now turn to the quantitative description of the system 
shown schematically in Fig.~\ref{fig:scheme}. 
Consider a one-dimensional propagation and interaction of a weak 
(quantum) probe field $\hat{\mc{E}}(z,t)$ with the atomic medium 
of linear density $\rho(z)$ and length $L$, taken as quantization length.
In the interaction picture \cite{MSSZ1997,PLDP2007}, the Hamiltonian 
of the system reads
\begin{eqnarray}
H/\hbar &=& -i \frac{c}{L} 
\int \! d z \, \hat{\mc{E}}^{\dag}(z) \, \partial_z \hat{\mc{E}}(z)
\nonumber \\ & & 
-  \int \! d z  \rho(z) \big[ \De \sig_{ee}(z) + (\De + \de) \sig_{rr} (z) 
\nonumber \\ & & \qquad \qquad \qquad
+ \big(\eta \hat{\mc{E}}(z) \sig_{eg} (z) + \mathrm{H.c.}\big) \big]  
\nonumber \\ & &
+ \int \! d z \rho(z) \big[ \sig_{re} (z) 
\sum_j^{n_s} D(z\bds{e}_z -\bds{r}_j) \sig^{(j)}_{-}
+ \mathrm{H.c.} \big] . \quad \; \label{eq:Ham}
\end{eqnarray}
Here the first term is the free Hamiltonian for the probe field 
$\hat{\mc{E}}(z) = \sum_k \hat{a}_k e^{ikz}$ propagating with velocity $c$. 
The probe field operators obey the commutation relations 
$[\hat{\mc{E}}(z),\hat{\mc{E}}(z')] 
= [\hat{\mc{E}}^{\dag}(z),\hat{\mc{E}}^{\dag}(z')] = 0$ and
$[\hat{\mc{E}}(z),\hat{\mc{E}}^{\dag}(z')] = L \de(z-z')$ which 
follow from the bosonic nature of operators $\hat{a}_k,\hat{a}^{\dag}_k$
for the individual longitudinal modes $k$. 
The second term of Eq.~(\ref{eq:Ham}) describes the atoms 
of the medium and their interaction with the probe field 
$\hat{\mc{E}}$ with the coupling strength 
$\eta = \wp_{ge} \sqrt{\om/(2 \hbar \eps_0 w^2 L)}$,
where $\wp_{ge}$ is the dipole matrix element of the transition
$\ket{g} \to \ket{e}$, $\om$ is the carrier frequency of the probe field,
$\eps_0$ is the vacuum permittivity, and $w \ll L$ is the probe 
field transverse width. We use the continuous atomic operators 
$\sig_{\mu\nu}(z) \equiv \frac{1}{N_z} \sum_i^{N_z} \ket{\mu}_i\bra{\nu}$
averaged over $N_z = \rho(z) \De z \gg 1$ atoms within 
a small interval $\De z$ around position $z$ \cite{EITrev2005}. 
These continuous operators obey the relations 
$\sig_{\mu\nu}(z) \sig_{\nu' \mu'}(z') = \sig_{\mu \mu'}(z) \de_{\nu \nu'} 
\de(z-z')/\rho(z)$. 
We work in the frame rotating with the optical $\om$ (probe field) 
and $\om_c$ (auxiliary coupling field) frequencies, and the microwave 
$\om_{ud}$ (spin-transition) frequency, as detailed in the next paragraph. 
Then the energy of the excited atomic level $\ket{e}$ is given by the 
detuning $\De = \om - \om_{eg}$ of the probe field from the transition 
resonance frequency $\om_{eg}$, and the energy of the Rydberg state 
$\ket{r}$ is defined via $\de = (\om_c + \om_{ud}) - \om_{re}$.  

The last term of Eq.~(\ref{eq:Ham}) describes the effective long-range 
interaction between the medium atoms and $n_s$ spins at positions $\bds{r}_j$.
These spins are represented by atoms with the Rydberg states $\ket{u}$ 
and $\ket{d}$ with the transition frequency $\om_{ud}$, and the spin 
lowering $\sig_-= \ket{d}\bra{u}$ and rising $\sig_+ = \ket{u}\bra{d}$ 
operators. The medium atoms and spins are coupled via the dipole-dipole 
interaction
\[
D_{as} = \frac{1}{4 \pi \eps_0 \hbar} 
\left[\frac{\bds{\wp}_{ri} \cdot \bds{\wp}_{du}}{|\bds{R}|^3} -
3 \frac{(\bds{\wp}_{ri} \cdot \bds{R}) (\bds{\wp}_{du} \cdot \bds{R})}
{|\bds{R}|^5}  \right] , 
\]
where $\bds{\wp}_{ri}$ is the dipole moment of the atomic transition 
$\ket{r} \lra \ket{i}$ between the Rydberg states $\ket{r}$ and $\ket{i}$,
$\bds{\wp}_{du}$ is the dipole moment of the spin transition 
$\ket{d} \lra \ket{u}$, and $\bds{R} \equiv  (z\bds{e}_z  -\bds{r})$ is 
the relative position vector between an atom at $z$ and a spin at $\bds{r}$.
To be specific, we consider a geometry of the system such that 
$\bds{\wp}_{ri} \parallel \bds{\wp}_{du} \perp \bds{R}$, i.e., $\bds{\wp}_{ri}$ 
and $\bds{\wp}_{du}$ are along the $y$ (quantization) axis, and assume 
that spin positions $\bds{r}_j$ are away from the $z$ axis, at $x_j > w$. 
Then the interaction $D_{as} = \frac{C_3}{|z\bds{e}_z  -\bds{r}_j|^3}$, with
$C_3 \equiv \frac{\wp_{ri} \wp_{du}}{4 \pi \eps_0 \hbar}$, is finite for all $z$.
The atomic excitation to the Rydberg state $\ket{r}$ is mediated 
by a non-resonant auxiliary coupling field of frequency $\omega_c$ 
which acts on the transition from the excited state $\ket{e}$ to the 
intermediate Rydberg state $\ket{i}$ with the Rabi frequency $\Omega_c$
and a large detuning $\De_c = \om_c - \om_{ie} = \om_{ri} - \om_{ud}$,
$|\De_c| \gg |D_{as}|, \Omega_c$. Upon adiabatic elimination of
the nonresonant state $\ket{i}$, we obtain the rate 
$D(z\bds{e}_z -\bds{r}_j) = \frac{C_3 \Om_c/\De_c}{|z\bds{e}_z  -\bds{r}_j|^3}$ 
of the effective atom-spin exchange interaction.
Due to negligible population of $\ket{i}$, we can then neglect 
the dipole-dipole interaction between the medium atoms, 
$D_{aa} \propto \frac{|\wp_{ri} \Om_c|^2}{\De_c^2}$, which requires that
$\left|\frac{\wp_{ri}}{\wp_{du}} \right| \ll \left|\frac{\De_c}{\Om_c}\right|$.

Adiabatic elimination of $\ket{i}$ leads also to the ac Stark shift 
$\frac{\Om_c^2}{\De_c}$ of level $\ket{e}$, which can be absorbed in 
the detuning $\De$, and to the dipole-dipole interaction induced 
shift $\de' = \frac{|D_{as}|^2}{-\De_c}$ of level $\ket{r}$, 
which should be added to $\de$.
Looking ahead to the DEIT resonance in the vicinity of $\De \simeq - \de$, 
we note that in order to be able to disregard the spatially varying 
shift $\de'$, we require it to be smaller than the DEIT linewidth 
$\frac{|D|^2}{|\ga_e +i \de|}$, where $\ga_e \geq \hlf \Ga_e$ is 
the relaxation rate of the $\sig_{ge}$ coherence \cite{EITrev2005}. 
This leads to the condition 
$\frac{|\ga_e +i \de|}{\Om_c} < \frac{\Om_c}{|\De_c|} \ll 1$, 
i.e., the Rabi frequency $\Om_c$ of the auxiliary field should be 
sufficiently larger than $\ga_e$ (setting $|\de| < \ga_e$ from now on), 
but still much smaller than $|\De_c|$.

We may assume that the $n_s$ spin-atoms are placed in a small volume of 
size $\De r \ll x_s/3$ at position $\bds{r}_s = x_s \bds{e}_x + z_s \bds{e}_z$ 
with $x_s \gg w$ and $z_s \simeq L/2$, such that the interaction strength
$D_{as}$ does not change appreciably within $\De r$. All the spin-atoms 
then couple symmetrically to the medium atoms, forming thereby an 
effective large spin $J=\hlf n_s$ of the Dicke model \cite{Dicke1954}
with the symmetric states $\ket{J,M_J}$ corresponding to $J + M_J$ 
atoms in state $\ket{u}$ and the remaining $J - M_J$ atoms in $\ket{d}$, 
where $M_J = -J,\ldots ,J$ is the ``magnetic'' (spin projection) quantum number.
With the collective spin-lowering
$\hat{J}_- \equiv \sum_j^{n_s} \sig_-^{(j)}$ and rising 
$\hat{J}_+ \equiv \sum_j^{n_s} \sig_+^{(j)}$ operators, the last
term of Hamiltonian~(\ref{eq:Ham}) can then be written as
$\int \! d z \rho(z) D(z) [ \sig_{re} (z) \hat{J}_{-} 
+ \hat{J}_{+} \sig_{er} (z)]$, where 
$D(z) \equiv D(z \bds{e}_z -\bds{r}_s)$. 
These operators obey standard spin-algebra relations:
$\hat{J}_- \ket{J,M_J} = \sqrt{(J+M_J)(J-M_J+1)} \ket{J,M_J-1}$,
$\hat{J}_+ \ket{J,M_J} = \sqrt{(J+M_J+1)(J-M_J)} \ket{J,M_J+1}$, 
$\hat{J}_z \ket{J,M_J} = M_J \ket{J,M_J}$, etc. 
Strictly speaking, we should also take into 
account the dipole-dipole exchange interactions between the spin-atoms, 
$\sum_{jj'} D_{ss}(\bds{r}_j -\bds{r}_{j'}) \sig_+^{(j)} \sig_-^{(j')}$, which 
generalizes the Dicke model to the Lipkin-Meshkov-Glick model \cite{LMG1965}. 
In the special case of infinite range interaction,
$D_{ss} = \mathrm{const} \; \forall \; j,j'$, we obtain the Hamiltonian
for the spin-$J$ as $H_J = h \hat{J}_z + D_{ss} (J^2 - \hat{J}_z^2)$, where 
$h$ is the effective magnetic field -- detuning, in the present context. 
For simplicity, we neglect the dispersive (van der Waals) interactions 
between the spin atoms. 
Below, our main concern is the case of at most a single spin-atom, $J=\hlf$,
but we will keep the notation generally applicable to any $J$,
assuming for simplicity negligible interactions between spins, 
leading to equidistant (or degenerate, for $h=0$) spectrum of $H_J$. 
Such a situation can in principle be realized for a few spin atoms 
arranged in certain geometric configurations, e.g., on a line tilted
by angle $\theta \simeq 54.7^{\circ}$ with respect to the direction of
the dipole moment vector $\bds{\wp}_{du}$ (the $y$ axis), since then
$D_{ss} \propto |\wp_{du}|^2 (1 - \cos^2 \theta) \simeq 0$.

Note finally that if, in the last term of Hamiltonian~(\ref{eq:Ham}),
we replace the dipolar exchange operator $D(z) \hat{J}_{-}$ 
(and its Hermite conjugate) with a c-number Rabi frequency of a 
classical driving field $\Om_d$, this Hamiltonian will describe
the usual EIT process \cite{Fleischhauer2000,EITrev2005,PLDP2007}.  
 
\subsection{Dynamics of the system}

From Hamiltonian~(\ref{eq:Ham}) we obtain the following Heisenberg 
equations for the relevant system operators:
\begin{eqnarray}
( \partial_t + c \partial_z ) \hat{\mc{E}}(z) &=& 
i \eta N \sig_{ge}(z) , \label{eq:Ep}  \\
\partial_t  \sig_{ge}(z) &=& (i \De - \ga_e) \sig_{ge}(z) 
\nonumber \\ & &
+ i \eta \hat{\mc{E}}(z) [\sig_{gg}(z) - \sig_{ee}(z)] 
\nonumber \\ & & 
- i D(z) \sig_{gr}(z) \hat{J}_+ + \hat{F}_{ge} , 
\label{eq:sigge} \\
\partial_t [\sig_{gr}(z) \hat{J}_+] &=& 
[i (\De + \de ) -\ga_r ] \sig_{gr}(z) \hat{J}_+ 
\nonumber \\ & &
- i \eta \hat{\mc{E}}(z) \sig_{er} (z) \hat{J}_+
\nonumber \\ & &
- i D(z) \sig_{ge}(z) \hat{J}_+ \hat{J}_- + \hat{F}_{gr},
\label{eq:siggrJp} 
\end{eqnarray}
where $N = \rho L\:$($\gg 1$) is the total number of medium atoms 
(assuming uniform density), and $\ga_e$ and $\ga_r\:$($\ll \ga_e$) 
are the atomic coherence relaxation rates 
(with $\hat{F}_{ge}$ and $\hat{F}_{gr}$ the associated Langevin noise operators),
while we ignore the decay of spins since Rydberg states $\ket{u}$ 
and $\ket{d}$ of spin-atoms are long-lived. 

\begin{figure}[t]
\centerline{\includegraphics[width=10cm]{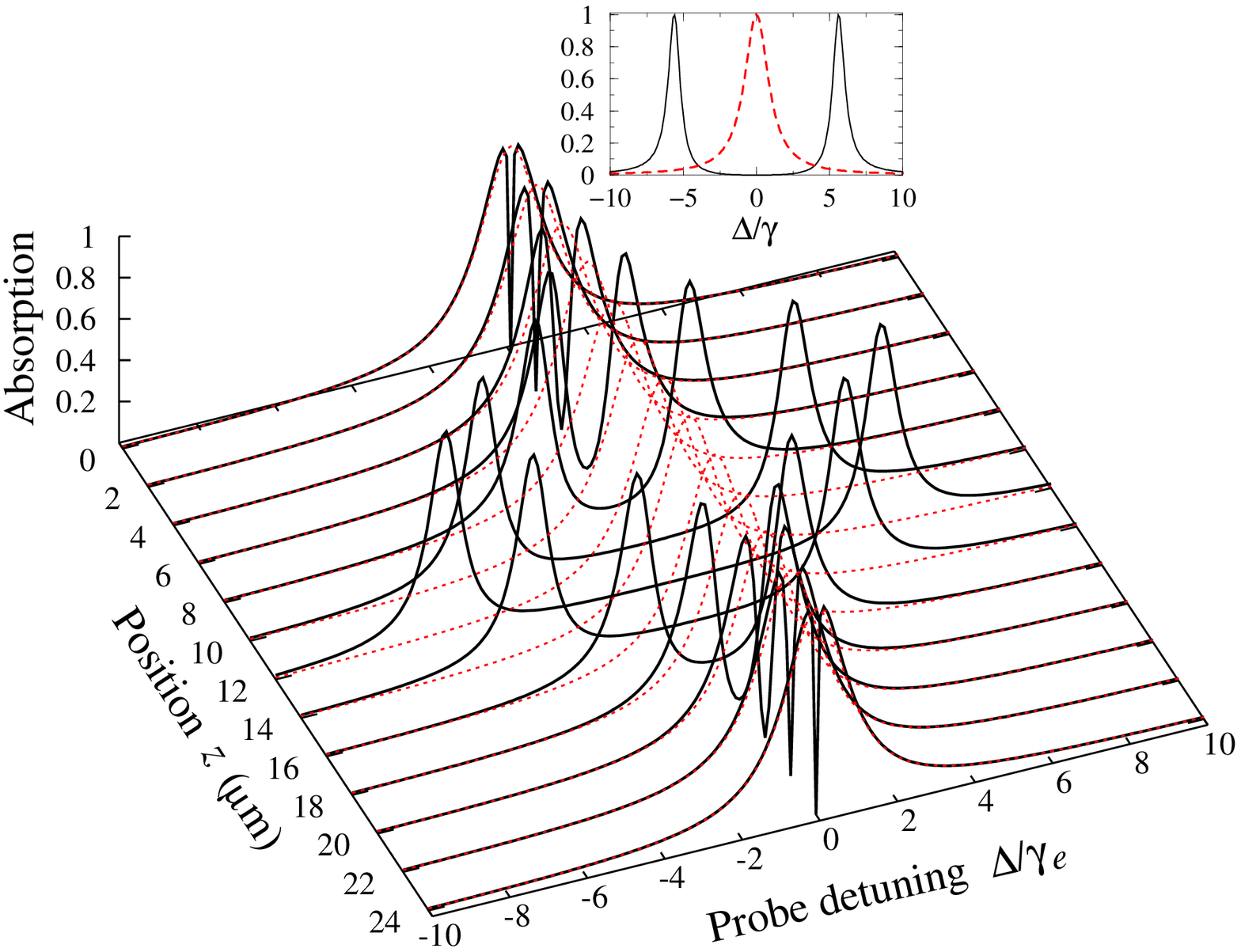}}
\caption{
Medium absorption $\frac{\om}{2c} \mathrm{Im} \expv{\hat \chi(z,\De)}$,
in units of the resonant absorption coefficient $\sigma_0 \bar{\rho}$,
as a function of position $z$ and probe frequency (detuning) $\De$, 
for $\expv{\hat{J}_+ \hat{J}_-} = 1$ (black solid lines) and 
$\expv{\hat{J}_+ \hat{J}_-} = 0$ (red dotted lines). Inset shows 
the DEIT  and TLA (position independent) absorption spectra at $z=12\:\mu$m.  
We set $\ga_r = 10^{-4} \ga_e$ and $\delta =0$, while $D(z)$ varies 
between $0.2 \ga_e$ (at $z = 0,L$) and $5.6 \ga_e$ (at $z = L/2$), 
as per parameters in Sec.~\ref{sec:expconcl}.} 
\label{fig:chiR}
\end{figure}

We consider adiabatic evolution of the system and drop the noise operators 
$\hat{F}$ since they do not contribute to the dynamics of the atomic and 
normally-ordered field operators \cite{Fleischhauer2000,EITrev2005}.
With all the medium atoms prepared initially in the ground state $\ket{g}$ 
and a weak input probe field ($n_p \ll N$), we can neglect the 
depletion of $\ket{g}$ and set $\sig_{gg} \simeq \mathds{1}$ and
$\sig_{ee},\sig_{er} \to 0$ in the above equations. From the stationary 
solution of Eqs.~(\ref{eq:sigge}), (\ref{eq:siggrJp}) we then obtain 
the atomic coherence $\sig_{ge}$. Substituting it in the field propagation 
Eq.~(\ref{eq:Ep}) without the time-derivative and comparing with
$\partial_z \hat{\mc{E}}=i\frac{\om}{2c} \chi \, \hat{\mc{E}}$ \cite{PLDP2007},
we obtain the medium susceptibility
\begin{equation}
\hat \chi(z,\De) = \frac{2}{\om} \frac{i \eta^2 N}
{\ga_e - i \De + \frac{|D(z)|^2 \hat{J}_+ \hat{J}_-}
{\ga_r - i (\De +\de)} } . \label{eq:chi}
\end{equation} 
Note that $2 \eta^2 N/\om = |\wp_{ge}|^2 \bar{\rho}/(\hbar\eps_0) 
= \frac{c}{\om}\sigma_0 \bar{\rho} \Ga_e$, where $\sigma_0 = 3\pi c^2/\om^2$ 
is the atomic resonant absorption cross-section assuming the 
(population) decay rate $\Ga_e = 2 \ga_e$ from state $\ket{e}$, 
and $\bar{\rho} = N/(w^2 L)$ is the volume density of atoms. 
Equation~(\ref{eq:chi}) has the form of the usual EIT susceptibility 
\cite{EITrev2005,Petrosyan2005,PLDP2007}, but with the square of the 
driving field Rabi frequency $|\Om_d|^2$ replaced by that of the 
space-dependent dipolar exchange operator, $|D(z)|^2 \hat{J}_+ \hat{J}_-$. 
In Fig.~\ref{fig:chiR} we show the imaginary part of medium susceptibility, 
Eq.~(\ref{eq:chi}), responsible for the probe absorption, as a function 
of probe frequency, at different spatial positions of the medium. 
In the absence of spin atom(s), $\expv{\hat{J}_+ \hat{J}_-} = 0$, 
the absorption spectrum is the usual Lorentzian of the TLA medium, 
while in the presence of a spin atom prepared in the spin-up 
state $\ket{u}$, such that $\expv{\hat{J}_+ \hat{J}_-} = 1$, 
the medium exhibits position-dependent EIT-like spectrum 
for a single probe photon (see below).

Using the expansion of $\hat{\chi}$ to first order in probe frequency 
around $\De$ \cite{PLDP2007}, we can now write the propagation equation 
for the probe pulse amplitude as
\begin{equation}
( \partial_t + \hat{v}_g \partial_z ) \hat{\mc{E}}(z,t) 
= i \frac{\om}{2} \hat{\chi} \, \hat{\mc{E}}(z,t),  \label{eq:EpvgChi}
\end{equation}
where $\hat{v}_g(z) = c [1 + \frac{\om}{2} 
\frac{\partial}{\partial \De } \mathrm{Re} \hat{\chi}(z,\De)]^{-1}$
is the group velocity \cite{EITrev2005,PLDP2007}, which, as the 
susceptibility, is $z$-dependent and operator-valued quantity. 
We are concerned with the dynamics of probe field with 
the carrier frequency near the DEIT (EIT) resonance $\De = - \de$, 
assuming the EIT-like condition 
$|\ga_e + i \de| \ga_r \ll |D(z)|^2 \; 
\forall \; z \in [0,L]$. 
The group velocity is then 
\begin{equation}
\hat{v}_g(z) = \frac{c}{ 1 
+ \frac{\eta^2 N}{|D(z)|^2 \hat{J}_+ \hat{J}_-}} \simeq 
c \frac{|D(z)|^2 \hat{J}_+ \hat{J}_-}{\eta^2 N} \ll c ,
\label{eq:vg}
\end{equation}
provided $\expv{\hat{J}_+ \hat{J}_-} \neq 0$ (see below), and assuming 
that the collective atom--field coupling $\eta^2 N$ is larger than 
the single atom--spin coupling $|D(z)|^2$ which remains finite 
even at $z \simeq L/2$ due to the spin position $x_s > w$. 
The propagation Eq.~(\ref{eq:EpvgChi}), supplemented with 
Eqs.~(\ref{eq:chi}) and (\ref{eq:vg}), is the central result of this paper.
Before we discuss its implications, however, we should establish 
the connection between the value of spin operator $\hat{J}_+ \hat{J}_-$
(for the given initial spin $J$) 
and the number of probe photons inside the medium $\hat{n}_p(t) = 
\frac{1}{L} \int_0^L d z \, \hat{\mc{E}}^{\dag}(z,t) \hat{\mc{E}}(z,t)$. 

Using Eq.~(\ref{eq:Ep}) and 
\begin{eqnarray}
\partial_t  \sig_{gg}(z) &=& i \eta \hat{\mc{E}}^{\dag}(z) \sig_{ge}(z)
+ \mathrm{H.c.} , 
\, \\
\partial_t  \hat{J}_z &=& \int \! d z \rho 
[i D(z) \sig_{re}(z) \hat{J}_- 
+ \mathrm{H.c.}] 
\nonumber \\
&=& - \int \! d z \rho \, \partial_t \sig_{rr}(z) , 
\end{eqnarray}
and taking into account that $\partial_t [\sig_{gg} + \sig_{rr}] =0$,
since under the DEIT (EIT) resonance and adiabatic evolution the excited 
state $\ket{e}$ is never populated, $\sig_{ee} =0$ \cite{EITrev2005,PLDP2007},
we obtain that $\partial_t [\hat{n}_p - \hat{J}_z] = 
\frac{c}{L} [\hat{\mc{E}}^{\dag}(0) \hat{\mc{E}}(0) 
- \hat{\mc{E}}^{\dag}(L) \hat{\mc{E}}(L)]$ is determined by the
difference of the flux of probe photons entering and leaving
the medium at $z=0$ and $z=L$, respectively. Next, from 
$\eta \hat{\mc{E}}(z) = D(z) \sig_{gr}(z) \hat{J}_+$ 
[cf. Eq.~(\ref{eq:sigge}) with $\sig_{ge},\sig_{ee} =0$] we have 
$\eta^2 \hat{n}_p = \frac{1}{L} \int_0^L d z \, |D(z)|^2 
\sig_{rr}(z) \hat{J}_- \hat{J}_+$.
We assume that the initial spin $J$ ($n_s = 2J$ spin atoms) is prepared
in state $\ket{J,J}$ (all spin-atoms in state $\ket{u}$). 
Using the equality $\hat{J}_- \hat{J}_+ = (J-\hat{J}_z)(J+\hat{J}_z+1)$,
after a little algebra we obtain the approximate expression
\begin{equation}
\eta^2 N \hat{n}_p \approx 2 J \bar{D}^2 (J - \hat{J}_z) , \label{eq:npsmall}
\end{equation}
where we used $\int_0^L \! d z \rho \, \sig_{rr}(z) = (J - \hat{J}_z)$ 
assuming that $\sig_{rr}(z)$ is a slowly varying function of $z$ 
in comparison to $|D(z)|^2$ with the mean value 
$\bar{D}^2 = \frac{1}{L} \int_0^L \! d z |D(z)|^2$. 
Recall that we positioned the spin $J$ such that
$\eta^2 N \gg |D(z\bds{e}_z -\bds{r}_s)|^2  \; \forall \; z \in [0,L]$. 
Equation~(\ref{eq:npsmall}) therefore indicates that inside 
the medium nearly all of probe photons are converted into 
the spin (de-)excitations, $\hat{n}_p \ll (J - \hat{J}_z)$. 
We then obtain that
\begin{eqnarray}
J - \hat{J}_z(t) & \simeq &\frac{c}{L} \int_0^t 
[\hat{\mc{E}}^{\dag}(0,t') \hat{\mc{E}}(0,t') 
- \hat{\mc{E}}^{\dag}(L,t') \hat{\mc{E}}(L,t')] d t' 
\nonumber \\
& \equiv & \hat{n}_{p,\mathrm{in}}(t) - \hat{n}_{p,\mathrm{out}}(t) .
\end{eqnarray}

We can now deduce the response of the medium to the incoming probe photons.
The $n_{p} = n_{p,\mathrm{in}} - n_{p,\mathrm{out}}$ photons, that already
entered the medium but not yet left it, are coherently converted
into the atomic Rydberg excitations $\ket{r}$ with simultaneous
flip of $n_{p}$ spin-atoms from state $\ket{u}$ to state $\ket{d}$,
corresponding to the spin state $\ket{J,J - n_{p}}$. Operator
$\hat{J}_+\hat{J}_-$ acting on that state leads to
$(n_s - n_{p})(n_{p} +1)$ which is non-zero if $n_s > n_{p}$.
Then the next probe photon entering the medium sees vanishing 
susceptibility, since in Eq.~(\ref{eq:chi}) the last term 
in denominator diverges under the DEIT (EIT) conditions. 
That $(n_p+1)$th probe photon propagates in the DEIT medium 
without absorption and with the $n_p$-dependent group velocity
$v_g^{(n_p+1)}(z) = c |D(z)|^2 (n_s - n_{p})(n_{p} +1)/(\eta^2 N)$
as per Eq.~(\ref{eq:vg}). We note parenthetically that if $n_s \gg n_{p}$ 
the large spin-$J$ behaves as a harmonic oscillator and the group velocity 
depends nearly linearly on $n_p$ -- a situation similar to VIT with 
$\Lambda$-atoms in a cavity \cite{Nikoghosyan2010}. 
On the other hand, if $n_s \leq n_{p}$, the susceptibility 
of Eq.~(\ref{eq:chi}) reduces to that of the resonant TLA 
medium ($|\De| < \ga_e$). Equation~(\ref{eq:EpvgChi}) then leads
to linear absorption of the incoming probe photon,
$\hat{\mc{E}}^{\dag}(z) \hat{\mc{E}}(z) = \hat{\mc{E}}^{\dag}(0) \hat{\mc{E}}(0)
e^{-\kappa z}$, with the (intensity) absorption coefficient 
$\kappa = 2 \sigma_0 \bar{\rho}$ \cite{PLDP2007}. 
Thus the DEIT medium behaves as a photon number filter, 
transmitting up to $n_p \leq n_s$ probe photons at a time, 
given the number $n_s$ of spin-atoms prepared in state $\ket{u}$.

Perhaps the most experimentally relevant and practically interesting 
situation of a single-photon filter or a transistor is realized for 
a single spin playing the role of a gate:
For $n_s = 0$ the medium is strongly absorbing for the incoming probe 
photons, with the optical depth $\mathrm{OD} = \kappa L$ which can be 
large enough in the medium of sufficient length $L$ (see below);
For $n_s = 1$ a single spin-atom in state $\ket{u}$ makes 
the medium transparent for one, and no more than one, probe photon
at a time.

\section{Experimental considerations and conclusions}
\label{sec:expconcl}

The system discussed above can be realized experimentally 
with currently available setups for Rydberg EIT with alkali atoms 
\cite{Peyronel2012,photboundsts,Pritchard2010,Hofmann2013,Maxwell2013,%
sphotswDuerr2014,sphotswHofferberth2014}. As a specific example, 
we may consider an ensemble of cold Rb atoms in an elongated trap 
of length $L \simeq 25 \: \mu$m. 
The ground $\ket{g}$ and excited $\ket{e}$ states of the medium atoms 
would correspond to suitable sublevels of the $5S_{1/2}$ and $5P_{1/2}$
(or $5P_{3/2}$) electronic states, with $\Ga_e \simeq 2\pi \times 6 \:$MHz, 
while the Rydberg states are $\ket{i} =  \ket{nS_{1/2},M_J =1/2}$ and 
$\ket{r} = \ket{nP_{3/2},M_J =1/2}$ with the principal quantum 
number $n \simeq 82$, and the quantization direction is taken along 
the $y$ axis ($\ket{i} \lra \ket{r}$ is a $\pi$-transition, $\Delta M_J =0$). 
The spin atom(s) can then be prepared by focused laser beam(s) in 
state $\ket{u} = \ket{(n'+1)S_{1/2},M_J =1/2}$ with strong dipole 
transition to state $\ket{d} = \ket{n'P_{1/2},M_J =1/2}$ ($\Delta M_J =0$).
A static external electric or magnetic field can lift the degeneracy 
of the Rydberg $M_J$ states of the spin and medium atoms to ensure that
only $\Delta M_J =0$ transitions are resonantly coupled via the 
dipole-dipole exchange interaction, in the presence of the auxiliary 
coupling field with the appropriate frequency $\om_c$ to satisfy the 
two-photon resonance condition $|\de| \ll \ga_e$ ($\de \simeq- \De$). 
With the quantum defects $\de_S = 3.131$ and $\de_P = 2.4565$ for 
the Rb $S$ and $P$ states \cite{RydAtoms}, we chose $n' = 86$ 
such that the transition $\ket{i} \to \ket{r}$ is appropriately detuned 
(by $\De_c \simeq 10 \Om_c \simeq 2\pi \times 0.44 \:$GHz)
from the $\ket{u} \to \ket{d}$ transition resonance. 
Calculation of the transition dipole moments $\wp_{ri}$ and $\wp_{du}$, 
involving the radial \cite{Kaulakys1995} and angular parts, leads to 
the coefficient $C_3 = 2\pi \times 10.8\:\mathrm{GHz} \: \mu$m$^3$. Then 
the DEIT linewidth $\de \om_{\mathrm{DEIT}} = |D(R)|^2/\ga_e$ of the medium
atoms can be large enough, $\de \om_{\mathrm{DEIT}} \gtrsim 2\pi \times 10^5\:$Hz 
at a distance $R \lesssim d_{\mathrm{t}} = 12.5 \:\mu$m from the spin atoms,
which permits the medium lengths $L \simeq 2 d_{\mathrm{t}}$. 
We take moderate atomic density $\bar{\rho} = 10^{12}\:$cm$^{-3}$ at which 
the mean interatomic separations is larger than the size of 
the $n \sim 80$ Rydberg electron orbit, so as to avoid excessive decoherence
due to electron collisions with the ground state atoms \cite{TPfau20134}. 
With $L \simeq 25 \:\mu$m we then obtain optical depth $\mathrm{OD} \simeq 7$, 
which would yield complete absorption of the probe photon(s) in the absence 
of DEIT, or saturation thereof by the previous $n_p =n_s$ photons. 
Decay and dephasing of Rydberg states of the medium and spin atoms, 
with the typical rate $\ga_r$ of several tens of kHz 
\cite{Peyronel2012,photboundsts,Pritchard2010,Petrosyan2011,Ates2011}, 
will degrade DEIT and lead to a small probability of absorption 
of the probe photon(s), 
$\int_0^L \! d z \frac{\om}{c} \mathrm{Im} \expv{\hat \chi(z)} 
\simeq 10^{-2}$, during propagation through the medium.
Several alternative choices of suitable atomic states and species 
are also possible. This will permit implementation of an efficient 
photon number switch as described above. 

Note finally that in the above analysis we have neglected van der Waals
interactions between the $\ket{r}$ state atoms. This simplification can 
be justified for a few probe photons simultaneously present in the medium, 
such that the mean distance between the photons is smaller than the van 
der Waals blockade distance $d_{\mathrm{b}} = \sqrt[6]{C_6/\de \om_{\mathrm{DEIT}}}$
\cite{Gorshkov2011,Petrosyan2011,Ates2011,Sevincli2011}. 
Due to the $R$ dependence of the DEIT linewidth,
$d_{\mathrm{b}} = \sqrt[6]{\frac{C_6 \ga_e \De_c^2}{C_3^2 \Om_c^2}} |R|$ varies 
between $\sim 3\:\mu$m in the center ($z \sim L/2$) and $\sim 10\:\mu$m at the
edges ($z \sim 0, L$) of the medium. Hence, even for many spin atoms 
$n_s > 1$, the number of probe photons in the medium is realistically 
limited to $n_p \leq 3$. Obviously, van der Waals interaction does not 
affect the performance of the single-photon switch, $n_s = 1$ or 0.

It would be interesting to consider an extended system with evenly 
distributed spin atoms prepared in one of the spin states with 
overlapping dipolar exchange field affecting the medium atoms. 
Then the density of probe photons that can propagate in the medium 
without attenuation will not exceed the density of spin atoms. 
Even more intriguing would be to explore simultaneous interaction 
of the Rydberg spin-atoms arranged in a chain-like 1D configuration 
with the medium atoms, and among themselves via resonant excitation 
(or hole) hoping. This may lead in bound states of the spin-flips 
(magnons) and propagating probe photons subject to DEIT. Developing 
an appropriate theoretical many-body description is a challenge worth 
pursuing as such systems could serve as viable quantum simulators
with quantum light fields and Rydberg atoms.

\begin{acknowledgments}
Useful discussion with Michael Fleischhauer, J\'ozsef Fort\'agh
and Klaus M\o lmer are gratefully acknowledged.
This work was supported in part by the H2020 FET Proactive project RySQ.
\end{acknowledgments}



\begin{thebibliography}{99}

\bibitem{RydAtoms}
T.F.~Gallagher, {\em Rydberg Atoms} (Cambridge University Press,
Cambridge, 1994)

\bibitem{rydQIrev}
M. Saffman, T.G. Walker, and K. M\o lmer,
{\it Quantum information with Rydberg atoms},
Rev. Mod. Phys. \textbf{82}, 2313 (2010)

\bibitem{rydDBrev}
D. Comparat and P. Pillet,
{\it Dipole blockade in a cold Rydberg atomic sample},
J. Opt. Soc. Am. B \textbf{27}, A208 (2010)

\bibitem{Jaksch2000}
D. Jaksch, J.I. Cirac, P. Zoller, S.L. Rolston, R. C\^ot\'e, and M.D. Lukin,
{\it Fast Quantum Gates for Neutral Atoms},
Phys. Rev. Lett. \textbf{85}, 2208 (2000)

\bibitem{Lukin2001}
M.D. Lukin, M. Fleischhauer, R. C\^ot\'e, L.M. Duan, D. Jaksch,
J.I. Cirac, and P. Zoller,
{\it Dipole Blockade and Quantum Information Processing 
in Mesoscopic Atomic Ensembles},
Phys. Rev. Lett. \textbf{87}, 037901 (2001)


\bibitem{Kiffner2012}
M. Kiffner, H. Park, W. Li, and T. F. Gallagher,
{\it Dipole-dipole-coupled double-Rydberg molecules},
Phys. Rev. A \textbf{86}, 031401(R) (2012)

\bibitem{Kiffner2013}
M. Kiffner, W. Li, and D. Jaksch,
{\it Three-Body Bound States in Dipole-Dipole Interacting Rydberg Atoms},
Phys. Rev. Lett. \textbf{111}, 233003 (2013)

\bibitem{Petrosyan2014}
D. Petrosyan and K. M\o lmer, 
{\it Binding potentials and interaction gates between microwave-dressed 
Rydberg atoms},
Phys. Rev. Lett. \textbf{113}, 123003 (2014)

\bibitem{Bettelli2013}
S. Bettelli, D. Maxwell, T. Fernholz, C. S. Adams, I. Lesanovsky, and C. Ates,
{\it Exciton dynamics in emergent Rydberg lattices},
Phys. Rev. A \textbf{88}, 043436 (2013)

\bibitem{Barredo2015}
D. Barredo, H. Labuhn, S. Ravets, T. Lahaye, A. Browaeys, and C. S. Adams,
{\it Coherent Excitation Transfer in a Spin Chain of Three Rydberg Atoms},
Phys. Rev. Lett. \textbf{114}, 113002 (2015)

\bibitem{Yu2016}
H. Yu and F. Robicheaux,
{\it Coherent dipole transport in a small grid of Rydberg atoms},
Phys. Rev. A \textbf{93}, 023618 (2016)

\bibitem{Whitlock2013}
G. G\"unter, H. Schempp, M. Robert-de-Saint-Vincent, V. Gavryusev, 
S. Helmrich, C. S. Hofmann, S. Whitlock, and M. Weidem\"uller,
{\it Observing the Dynamics of Dipole-Mediated Energy Transport},
Science \textbf{342}, 954 (2013)

\bibitem{Whitlock2015}
D.W. Sch\"onleber, A. Eisfeld, M. Genkin, S. Whitlock, and S. W\"uster,
{\it Quantum Simulation of Energy Transport with Embedded Rydberg Aggregates},
Phys. Rev. Lett. \textbf{114}, 123005 (2015);
H. Schempp, G. G\"unter, S. W\"uster, M. Weidem\"uller, and S. Whitlock,
{\it Correlated Exciton Transport in Rydberg-Dressed-Atom Spin Chains},
Phys. Rev. Lett. \textbf{115}, 093002 (2015)


\bibitem{Fleischhauer2000}
M. Fleischhauer and M. D. Lukin, 
{\it Dark-State Polaritons in Electromagnetically Induced Transparency},
Phys. Rev. Lett. \textbf{84}, 5094 (2000);
M. Fleischhauer and M. D. Lukin,
{\it Quantum memory for photons: Dark-state polaritons},
Phys. Rev. A \textbf{65}, 022314 (2002)


\bibitem{EITrev2005}
M. Fleischhauer, A. Imamoglu, and J. P. Marangos,
{\it Electromagnetically induced transparency: Optics in coherent media},
Rev. Mod. Phys. \textbf{77}, 633 (2005)



\bibitem{Field1993}
J. E. Field, 
{\it Vacuum-Rabi-splitting-induced transparency},
Phys. Rev. A \textbf{47}, 5064 (1993)

\bibitem{Tanji-Suzuki2011}
H. Tanji-Suzuki, W. Chen, R. Landig, J. Simon, and V. Vuletic,
{\it Vacuum-Induced Transparency},
Science \textbf{333}, 1266 (2011)

\bibitem{Nikoghosyan2010}
G. Nikoghosyan and M. Fleischhauer,
{\it Photon-Number Selective Group Delay in Cavity Induced Transparency},
Phys. Rev. Lett. \textbf{105}, 013601 (2010);
N. Lauk and M. Fleischhauer, 
{\it Number-state filter for pulses of light},
Phys. Rev. A \textbf{93}, 063818 (2016)




\bibitem{Friedler2005}
I. Friedler, D. Petrosyan, M. Fleischhauer and G. Kurizki, 
{\it Long-range interactions and entanglement of slow single-photon pulses},
Phys. Rev. A \textbf{72}, 043803 (2005);
E. Shahmoon, G. Kurizki, M. Fleischhauer and D. Petrosyan, 
{\it Strongly interacting photons in hollow-core waveguides},
Phys. Rev. A \textbf{83}, 033806 (2011)

\bibitem{He2011}
B. He, A. MacRae, Y. Han, A. I. Lvovsky, and C. Simon,
Phys. Rev. A \textbf{83}, 022312 (2011);
B. He, A. V. Sharypov, J. Sheng, C. Simon, and M. Xiao,
Phys. Rev. Lett. \textbf{112}, 133606 (2014)

\bibitem{Parigi2012}
V. Parigi, E. Bimbard, J. Stanojevic, A. J. Hilliard, F. Nogrette, 
R. Tualle-Brouri, A. Ourjoumtsev, and P. Grangier,
{\it Observation and Measurement of Interaction-Induced Dispersive 
Optical Nonlinearities in an Ensemble of Cold Rydberg Atoms},
Phys. Rev. Lett. \textbf{109}, 233602 (2012)

\bibitem{Tiarks2016}
D. Tiarks, S. Schmidt, G. Rempe and S. D\"urr,
{\it Optical $\pi$ phase shift created with a single-photon pulse},
Sci. Adv. \textbf{2}, e1600036 (2016)

\bibitem{Paredes2014}
D. Paredes-Barato and C. S. Adams,
{\it All-Optical Quantum Information Processing Using Rydberg Gates},
Phys. Rev. Lett. \textbf{112}, 040501 (2014)


\bibitem{Gorshkov2011}
A. V. Gorshkov, J. Otterbach, M. Fleischhauer, T. Pohl, and M. D. Lukin,
{\it Photon-Photon Interactions via Rydberg Blockade},
Phys. Rev. Lett. \textbf{107}, 133602 (2011)


\bibitem{Peyronel2012}
T. Peyronel, O. Firstenberg, Q.-Y. Liang, S. Hofferberth, A. V. Gorshkov,
T. Pohl, M. D. Lukin, and V. Vuleti\'c,
{\it Quantum nonlinear optics with single photons 
enabled by strongly interacting atoms},
Nature \textbf{488}, 57 (2012)

\bibitem{photboundsts}
O. Firstenberg,	T. Peyronel, Q.-Y. Liang, A. V. Gorshkov, M. D. Lukin,
and V. Vuleti\'c,
{\it Attractive photons in a quantum nonlinear medium},
Nature \textbf{502}, 71 (2013)

\bibitem{Pritchard2010}
J. D. Pritchard, D. Maxwell, A. Gauguet, K. J. Weatherill, 
M. P. A. Jones, and C. S. Adams,
{\it Cooperative Atom-Light Interaction in a Blockaded Rydberg Ensemble},
Phys. Rev. Lett. \textbf{105}, 193603 (2010)

\bibitem{Petrosyan2011}
D. Petrosyan, J. Otterbach and M. Fleischhauer,
{\it Electromagnetically induced transparency with Rydberg atoms},
Phys. Rev. Lett. \textbf{107}, 213601 (2011)

\bibitem{Ates2011}
C. Ates, S. Sevincli, and T. Pohl,
{\it Electromagnetically induced transparency in strongly interacting 
Rydberg gases},
Phys. Rev. A \textbf{83}, 041802(R) (2011)

\bibitem{Sevincli2011}
S. Sevincli, N. Henkel, C. Ates, and T. Pohl,
{\it Nonlocal Nonlinear Optics in Cold Rydberg Gases},
Phys. Rev. Lett. \textbf{107}, 153001 (2011)

\bibitem{Hofmann2013}
C. S. Hofmann, G. G\"unter, H. Schempp, M. Robert-de-Saint-Vincent, 
M. G\"arttner, J. Evers, S. Whitlock, and M. Weidem\"uller,
{\it Sub-Poissonian Statistics of Rydberg-Interacting Dark-State Polaritons},
Phys. Rev. Lett. \textbf{110}, 203601 (2013)



\bibitem{Kuzmich13}
Y.O. Dudin and A. Kuzmich,
{\it Strongly Interacting Rydberg Excitations of a Cold Atomic Gas},
Science \textbf{336}, 887 (2012)

\bibitem{Otterbach2013}
J. Otterbach, M. Moos, D. Muth, and M. Fleischhauer,
{\it Wigner crystallization of photons in cold Rydberg ensembles},
Phys. Rev. Lett. \textbf{111}, 113001 (2013);
M. Moos, M. H\"oning, R. Unanyan, and M. Fleischhauer,
{\it Many-body physics of Rydberg dark-state polaritons 
in the strongly interacting regime},
Phys. Rev. A \textbf{92}, 053846 (2015)

\bibitem{Distante2016}
E. Distante, A. Padron-Brito, M. Cristiani, D. Paredes-Barato, 
and H. de Riedmatten,
{\it Storage Enhanced Nonlinearities in a Cold Atomic Rydberg Ensemble}
Phys. Rev. Lett. \textbf{117}, 113001 (2016)



\bibitem{Li2014}
W. Li, D. Viscor, S. Hofferberth, and I. Lesanovsky,
{\it Electromagnetically Induced Transparency in an Entangled Medium},
Phys. Rev. Lett. \textbf{112}, 243601 (2014).


\bibitem{Maxwell2013}
D. Maxwell, D. J. Szwer, D. Paredes-Barato, H. Busche, J. D. Pritchard, 
A. Gauguet, K. J. Weatherill, M. P. A. Jones, and C. S. Adams,
{\it Storage and Control of Optical Photons Using Rydberg Polaritons},
Phys. Rev. Lett. \textbf{110}, 103001 (2013)

\bibitem{sphotswDuerr2014}
S. Baur, D. Tiarks, G. Rempe, and S. D\"urr, 
{\it Single-Photon Switch Based on Rydberg Blockade},
Phys. Rev. Lett. \textbf{112}, 073901 (2014);
D. Tiarks, S. Baur, K. Schneider, S. D\"urr, and G. Rempe,
{\it Single-Photon Transistor Using a F\"orster Resonance},
Phys. Rev. Lett. \textbf{113}, 053602 (2014)

\bibitem{sphotswHofferberth2014}
H. Gorniaczyk, C. Tresp, J. Schmidt, H. Fedder, and S. Hofferberth,
{\it Single-Photon Transistor Mediated by Interstate Rydberg Interactions},
Phys. Rev. Lett. \textbf{113}, 053601 (2014);
H. Gorniaczyk, C. Tresp, P. Bienias, A. Paris-Mandoki, W. Li, 
I. Mirgorodskiy, H. P. B\"uchler, I. Lesanovsky, and S. Hofferberth,
{\it Enhancement of Rydberg-mediated single-photon nonlinearities 
by electrically tuned F\"orster resonances},
Nature Commun. \textbf{7}, 12480 (2016)

\bibitem{PohlRev2016}
C. Murray and T. Pohl,
{\it Quantum and Nonlinear Optics in Strongly Interacting Atomic Ensembles},
Adv. Atom. Mol. Opt. Phys. \textbf{65}, 321 (2016).

\bibitem{FirstenbergRev2016}
O. Firstenberg, C. S. Adams, S. Hofferberth,
{\it Nonlinear quantum optics mediated by Rydberg interactions},
J. Phys. B \textbf{49}, 152003 (2016)


\bibitem{TPfau20134}
J. B. Balewski,	A. T. Krupp, A. Gaj, D. Peter, H. P. B\"uchler, 
R. L\"ow, S. Hofferberth, and T. Pfau,
{\it Coupling a single electron to a Bose–Einstein condensate},
Nature \textbf{502}, 664 (2013);
A. Gaj,	A. T. Krupp, J. B. Balewski, R. L\"ow, S. Hofferberth, and T. Pfau,
{\it From molecular spectra to a density shift in dense Rydberg gases},
Nat. Commun. \textbf{5}, 4546 (2014)



\bibitem{MSSZ1997}
M. O. Scully and M. S. Zubairy,
{\it Quantum Optics},
(Cambridge University Press, Cambridge, UK, 1997)

\bibitem{PLDP2007}
P. Lambropoulos and D. Petrosyan,
{\it Fundamentals of Quantum Optics and Quantum Information},
(Springer, Berlin, 2007)


\bibitem{Dicke1954}
R. H. Dicke, 
{\it Coherence in Spontaneous Radiation Processes},
Phys. Rev. \textbf{93}, 99 (1954)

\bibitem{LMG1965}
H. J. Lipkin, N. Meshkov, and A. J. Glick, 
{\it Validity of many-body approximation methods for a solvable model: 
(I). Exact solutions and perturbation theory},
Nucl. Phys. \textbf{62}, 188 (1965)



\bibitem{Petrosyan2005}
D. Petrosyan, 
{\it Towards deterministic optical quantum computation 
with coherently driven atomic ensembles},
J. Opt. B \textbf{7}, S141 (2005)

\bibitem{Kaulakys1995}
B. Kaulakys,
{\it Consistent analytical approach for the quasi-classical 
radial dipole matrix elements},
J . Phys. B \textbf{28}, 4963 (1995)

\end{thebibliography}
\end{document}